\documentclass[prb,twocolumn]{revtex4}
\usepackage{graphicx}
\usepackage{dcolumn}
\usepackage{bm}

\begin{document}

\preprint{ATB-1}

\title{The magnetic state of Yb in Kondo-lattice
YbNi$_{2}$B$_{2}$C}

\author{A. T. Boothroyd}
\email{a.boothroyd1@physics.ox.ac.uk}
\homepage{http://xray.physics.ox.ac.uk/Boothroyd}\affiliation{
Department of Physics, Oxford University, Oxford, OX1 3PU, United
Kingdom }
\author{J. P. Barratt}\affiliation{ Department of Physics, Oxford University,
Oxford, OX1 3PU, United Kingdom }\affiliation{ Institut
Laue-Langevin, BP 156, 38042 Grenoble Cedex 9, France}
\author{P. Bonville}\affiliation{
Centre d'Etudes de Saclay, DSM-DRECAM, Service de Physique de
l'Etat Condens\'{e}, 91191 Gif-sur-Yvette Cedex, France }
\author{P. C. Canfield}\affiliation{
Iowa State University and Ames Laboratory, Ames, Iowa 50011,
U.S.A. }
\author{A. Murani}\affiliation{
Institut Laue-Langevin, BP 156, 38042 Grenoble Cedex 9, France }
\author{A. R. Wildes}\affiliation{
Institut Laue-Langevin, BP 156, 38042 Grenoble Cedex 9, France }
\author{R. I. Bewley}\affiliation{
ISIS Facility, Rutherford Appleton Laboratory, Chilton, Didcot,
OX11 0QX, United Kingdom }

\date{\today}

\begin{abstract}
We report neutron scattering experiments performed to investigate
the dynamic magnetic properties of the Kondo-lattice compound
YbNi$_2$B$_2$C. The spectrum of magnetic excitations is found to
be broad, extending up to at least 150\,meV, and contains
inelastic peaks centred near 18\,meV and 43\,meV. At low energies
we observe quasielastic scattering with a width $\Gamma =
2.1$\,meV.  The results suggest a Yb$^{3+}$ ground state with
predominantly localized 4$f$ electrons subject to (i) a
crystalline electric field (CEF) potential, and (ii) a Kondo
interaction, which at low temperatures is about an order of
magnitude smaller than the CEF interaction. From an analysis of
the dynamic magnetic response we conclude that the crystalline
electric field acting on the Yb ions has a similar anisotropy to
that in other $R$Ni$_2$B$_2$C compounds, but is uniformly enhanced
by almost a factor of 2. The static and dynamic magnetic
properties of YbNi$_2$B$_2$C are found to be reconciled quite well
by means of an approximation scheme to the Anderson impurity
model, and this procedure also indicates that the effective Kondo
interaction varies with temperature due to the crystal field
splitting. We discuss the nature of the correlated-electron ground
state of YbNi$_2$B$_2$C based on these and other experimental
results, and suggest that this compound might be close to a
quantum critical point on the non-magnetic side.
\end{abstract}

\maketitle

\section{\label{sec:intro}Introduction}

One of the delights of $f$-electron metals is their tendency to
show beautiful electronic ordering phenomena at low temperatures.
Typically one can expect to find long-range magnetic ordering,
superconductivity, heavy-electron behaviour, or quantum critical
effects, and sometimes two or more of these in the same material.
The character of the ground state is often governed by several
different microscopic interactions of comparable strength.

The diversity of possible electronic phases is exemplified by the
$R$Ni$_2$B$_2$C family, $R$ being Sc, Y, most lanthanides, and
some actinides.\cite{Nagarajan-PRL-1994, Cava-Nature-1994,
Muller-RPP-2001, Canfield-PhysicsTD-1998} Most members of this
family exhibit either magnetic ordering or superconductivity, and
for $R$ = Dy, Ho, Er and Tm there exists a temperature range over
which superconductivity and magnetic order coexist. These four
members of the family have been the subject of extensive
investigations into the interplay between magnetism and
superconductivity. The materials with $R$ = Sc, Y, Lu, Th and
possibly Ce are non-magnetic superconductors,\cite{Lai-PRB-1995,
Massalami-PhysicaC-1998} whereas LaNi$_2$B$_2$C is a conventional
metal. Among the remaining members of the series, only
YbNi$_2$B$_2$C displays neither magnetic ordering nor
superconductivity. This compound is metallic with enhanced
electronic transport and thermodynamic coefficients at low
temperatures, and for this reason it has been classified a heavy
fermion system.\cite{Dhar-SSC-1996, Yatskar-PRB-1996}

This paper addresses the nature of the ground state of
YbNi$_2$B$_2$C. Since it is stoichiometric, YbNi$_2$B$_2$C is one
of a rather small number of known Yb Kondo-lattice compounds. Its
properties are then expected to be strongly influenced by local
Kondo singlet formation, with competition from long-range magnetic
ordering of localized 4$f$ moments coupled by the RKKY exchange
interaction. Given the right energy balance, Kondo-lattice
compounds can exhibit a coherent Fermi liquid phase at low
temperatures with strongly renormalized quasiparticles, or `heavy
fermions'. In the heavy fermion state an abundance of low-energy
spin fluctuations mediate the interactions between the
quasiparticles and strongly influence the physical properties. An
additional factor is the crystalline electric field (CEF), which
by splitting the 4$f$ levels dictates the symmetry of the 4$f$
ground state and makes the effective 4$f$ magnetic moment
temperature dependent.

Evidence for a coherent heavy fermion state in YbNi$_2$B$_2$C has
been found in several different properties.\cite{Dhar-SSC-1996,
Yatskar-PRB-1996} The electronic heat capacity has a peak near
8\,K, and decreases below 5\,K with a large slope corresponding to
a Sommerfeld coefficient $\gamma \approx
0.5$\,J\,mol$^{-1}$K$^{-2}$.\cite{Yatskar-PRB-1996} This value of
$\gamma$ translates into a single-impurity Kondo temperature
$T_{\rm K}\approx10$\,K. The resistivity decreases steadily below
room temperature and then more sharply below a temperature in the
range 10--40\,K depending on the annealing treatment of the
sample\cite{Avila-cond-mat-2002}. At temperatures below
$\sim$1.5\,K the temperature dependence of the resistivity
exhibits a positive curvature reminiscent of the $T^2$ variation
characteristic of a strongly-correlated Fermi liquid. The
temperature dependence of the spin-lattice relaxation rate
measured by $^{11}{\rm B}$ nuclear magnetic
resonance\cite{Sala-PRB-1997} shows a transition from local moment
relaxation at high temperatures to Korringa-like behaviour below
5\,K suggestive of a Fermi liquid with a high density of states.
Finally, no magnetic ordering has been observed in YbNi$_2$B$_2$C
down to 0.023\,K,\cite{Bonville-EPJ-1999} a temperature much less
than the Yb magnetic ordering temperature of 0.4\,K predicted by
de Gennes scaling from the heavy $R$ ions in $R$Ni$_2$B$_2$C, and
nearly four orders below the Weiss temperature $\theta_{\rm
p}\approx -100$\,K derived from a Curie-Weiss law fit to the
high-temperature susceptibility. These observations suggest the
existence of a strong antiferromagnetic interaction between Yb and
conduction band states, which causes a screening of the Yb 4$f$
moment at low temperatures by the Kondo effect.

Here we describe neutron inelastic scattering measurements of the
magnetic excitation spectrum of YbNi$_2$B$_2$C in the 0--200\,meV
energy range. This technique directly measures the spin
fluctuation spectrum, and allows us to determine the Kondo, CEF
and RKKY energy scales quantitatively. Our main findings are (i)
that the CEF potential is larger than either the Kondo or RKKY
interactions, and (ii) that the magnetic excitation spectra and
other bulk magnetic properties can be quite well understood in
terms of a model containing CEF and Kondo interactions provided
that the latter is allowed to increase with temperature. A
preliminary account of our experimental data was reported in Ref.
\onlinecite{Boothroyd-NATOARW-2001}. Neutron spectroscopic data
have also been published by Sierks {it et
al}.,\cite{Sierks-PhysicaB-1999} and though our conclusions are
partly in accord with those of Sierks {\it et al}. there are also
differences which we will seek to explain later.

\section{\label{sec:expt}Sample preparation}

Experiments were performed on single-crystalline and
polycrystalline YbNi$_2$B$_2$C. All the samples were prepared with
boron enriched to 99.5\% $^{11}\rm B$ to reduce neutron absorption
by the $^{10}\rm B$ isotope present in natural boron.
Single-crystalline LuNi$_2$B$_2$C and polycrystalline
Y$_{0.5}$Lu$_{0.5}$Ni$_2$B$_2$C were used as non-magnetic
reference samples.

The single crystals were grown at Ames Laboratory by the
high-temperature Ni$_2$B flux method.\cite{Xu-PhysicaB-1994,
Cho-PRB-1995} In the case of YbNi$_2$B$_2$C, excess Yb was added
to compensate for the loss due to evaporation. The crystals were
plate-like with typical dimensions $4 \times 4 \times
0.3$\,mm$^3$, and the mass of the largest crystal was 0.25\,g. For
the neutron experiments we prepared a mosaic of 40 crystals with a
total mass of approximately 1\,g. The crystals were co-aligned to
within 5$^\circ$ and glued onto a thin sheet of aluminium. A
similar mosaic of LuNi$_2$B$_2$C crystals was used to estimate the
non-magnetic signal.

Polycrystalline YbNi$_2$B$_2$C was prepared by a solid-state
reaction method similar to that described by Dhar {\it et
al},\cite{Dhar-SSC-1996}
and polycrystalline Y$_{0.5}$Lu$_{0.5}$Ni$_2$B$_2$C was prepared
by the standard arc-melting technique under flowing argon. X-ray
powder diffraction was used to check for impurities, and the
YbNi$_2$B$_2$C sample was found to contain a measurable amount of
Yb$_2$O$_3$. Subsequently, a quantitative analysis of the
composition of this sample was carried out by multi-phase
refinement of neutron powder diffraction data collected on the D1b
diffractometer at the Institut Laue-Langevin (ILL). The amount of
Yb$_2$O$_3$ impurity was found to be 9\% by mass. As described
below, we corrected for the spurious signal produced by this
impurity by performing measurements on commercially-obtained
Yb$_2$O$_3$ powder (99.9\% purity) under identical conditions as
used for YbNi$_2$B$_2$C, with Y$_2$O$_3$ powder as the
non-magnetic reference.

\section{\label{sec:inelastic}Magnetic excitation spectrum}

\subsection{\label{subsec:inel_theory}Neutron inelastic scattering cross-section}

The quantity measured by neutron inelastic scattering is the
differential cross-section per unit solid angle $\Omega$ and
scattered neutron energy $E_{\rm f}$, given by
\begin{equation}
\frac{{\rm d}^2\sigma}{{\rm d}\Omega{\rm d}E_{\rm f}} =
\frac{k_{\rm f}}{k_{\rm i}}\,S({\bf
Q},\omega).\label{eq:xs_response_fn}
\end{equation}
Here, $k_{\rm i}$ and $k_{\rm f}$ are the incident and scattered
neutron wavevectors, $S({\bf Q},\omega)$ is the response function
of the sample, ${\bf Q} = {\bf k}_{\rm i} - {\bf k}_{\rm f}$ is
the scattering vector, and $\hbar\omega = E_{\rm i} - E_{\rm f}$
is the energy transferred from the neutron to the sample. From
linear response theory it can be shown that the dynamical part of
the response function for a paramagnetic ion is given in the
dipole approximation by\cite{Lovesey}
\begin{eqnarray}
\tilde{S}({\bf Q},\omega)&=& \frac{\left(\gamma r_0\right)^2}{4
\mu_0 \mu_{\text B}^2} \exp\{-2W({\bf
Q})\}|f({\bf Q})|^2\,\omega\{1+n(\omega)\}\nonumber\\
& & \times \sum_{\alpha}(1-\hat{Q}_{\alpha}^2 )
\chi^{\alpha\alpha}F^{\alpha \alpha}(\omega),
\label{eq:dynamic_res_fn_crystal}
\end{eqnarray}
where $\gamma=-1.913$, $r_0=2.818\times 10^{-15}$\,m is the
classical electron radius, $\exp\{-2W({\bf Q})\}$ is the
Debye-Waller factor, $|f({\bf Q})|^2$ is the squared modulus of
the magnetic form factor, $n(\omega)=1/\{\exp(\hbar\omega/k_{\rm
B}T)-1\}$ is the Planck distribution, $\hat{Q}_{\alpha}$ is the
$\alpha$ ($ = x, y, z$) component of the unit scattering vector,
$\chi^{\alpha\alpha}$ is a leading-diagonal element of the static
single-ion susceptibility tensor\cite{footnote1}, and $F^{\alpha
\alpha}(\omega)$ is a spectral-weight function with unit
normalization:
\begin{equation}
\int_{-\infty}^{\infty}F^{\alpha \alpha}(\omega)\,{\rm d}\omega =
1.\label{eq:spec_wt_fn_norm_ch3}
\end{equation}
In the case of a polycrystalline sample, Eq.\
(\ref{eq:dynamic_res_fn_crystal}) must be averaged over all
orientations. If, in addition, we extrapolate the scattering to
zero $\bf Q$ then the expression for the dynamic part of the
response function becomes
\begin{eqnarray}
\tilde{S}(0,\omega)&=& \frac{\left(\gamma r_0\right)^2}{4 \mu_0
\mu_{\text B}^2}\,\omega\{1+n(\omega)\} 2\chi_{\text av}F(\omega),
\label{eq:dynamic_res_fn_powder}
\end{eqnarray}
where $\chi_{\text av} =
\frac{1}{3}(\chi^{xx}+\chi^{yy}+\chi^{zz})$ is the powder-averaged
static susceptibility.

\subsection{\label{subsec:inel_expt}Experimental details}

We employed four different neutron spectrometers to study the
magnetic excitation spectrum. The polycrystalline samples were
measured on the high energy transfer (HET) chopper spectrometer at
the ISIS spallation neutron source and on the IN5 chopper
spectrometer at the Institut Laue-Langevin (ILL). These
experiments probed the energy ranges 5--200\,meV and 0.3--2.5\,meV
respectively. The single-crystal samples were measured on the IN14
triple-axis spectrometer at the ILL and on the IN6 time-of-flight
spectrometer, also at the ILL. These latter experiments provided
information on the $\bf Q$ dependence of the low-energy
excitations in the energy ranges 0.2--6\,meV (IN14) and
0.2--4.5\,meV (IN6). All four spectrometers were equipped with a
variable-temperature liquid helium cryostat allowing measurements
to be made as a function of temperature. Experimental details
specific to each spectrometer are as follows:

(1) HET. Data were collected in three runs, with neutrons of
incident energy 35\,meV, 75\,meV and 250\,meV respectively.
Spectra recorded in banks of detectors distributed around the
incident beam direction were averaged.  The mass of
polycrystalline YbNi$_2$B$_2$C in the beam was approximately
13\,g. To provide an estimate of the non-magnetic background
scattering we also measured spectra at the same incident energies
from a similar mass of polycrystalline
Y$_{0.5}$Lu$_{0.5}$Ni$_2$B$_2$C. This composition was chosen
because it has the same cross-section for neutron absorption as
YbNi$_2$B$_2$C.

(2) IN5. Measurements were made with the incident energy fixed at
3.1\,meV. The data did not show any detectable $\bf Q$ dependence
apart from a very slow reduction in intensity with $|\bf Q|$ for a
given energy consistent with the variation of the magnetic form
factor, and so the counts recorded in the whole detector bank
(extending from $\sim$15$^\circ$ to $\sim$130$^\circ$ in
scattering angle) were averaged to improve statistics.

(3) IN14. Spectra were recorded by scanning the incident neutron
energy with a fixed final energy of $E_{\rm f}=4.7$\,meV. The
incident and final energies were selected by Bragg reflection from
arrays of pyrolytic graphite crystals. A beryllium filter was
placed immediately after the sample to suppress higher-order
harmonics in the scattered beam. Two settings of the crystals were
used, giving access to the $(h, h, l)$ and $(h, 0, l)$ planes in
reciprocal space.

(4) IN6. The incident neutron energy was fixed at either 3.1\,meV
or 4.9\,meV. As on IN5, we averaged the neutron counts recorded in
the whole detector ($\sim$10$^\circ$ to $\sim$115$^\circ$). This
means that the recorded spectra correspond to an average over a
range of $\bf Q$. We chose two orientations of the crystal
relative to the incident beam so that in one case the average $\bf
Q$ was approximately parallel to the $a$ axis of the crystal, and
in the other it was approximately parallel to the $c$ axis.

On all the time-of-flight spectrometers (HET, IN5 and IN6) the
scattering from a standard vanadium sample was used to normalize
the data in different detector banks and to convert the spectra
into units of absolute scattering cross-section.

\subsection{\label{subsec:inel_res}Results}

\begin{figure}
\includegraphics{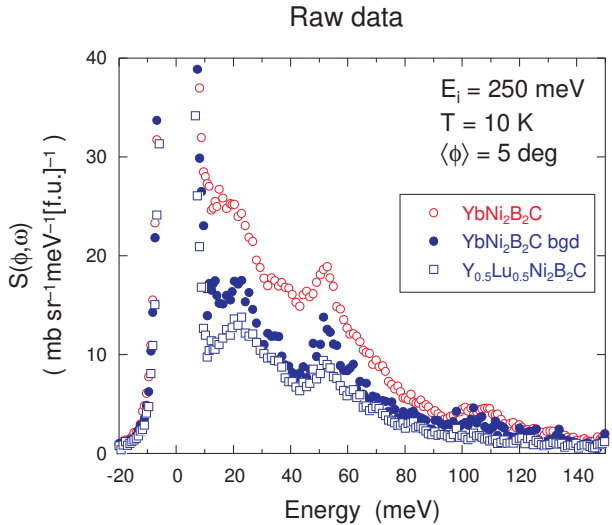}
\caption{Raw neutron scattering energy spectra obtained from
polycrystalline samples of YbNi$_2$B$_2$C and
Y$_{0.5}$Lu$_{0.5}$Ni$_2$B$_2$C on the HET time-of-flight
spectrometer. The incident neutron energy was 250\,meV, and the
data shown were recorded in the HET low-angle detector bank and
normalized per formula unit (f.u.) of YbNi$_2$B$_2$C. The
YbNi$_2$B$_2$C non-magnetic background scattering (filled circles)
was estimated by the method described in the text. \label{fig:1} }
\end{figure}

We begin with the high-energy data collected on HET.
Figure\,\ref{fig:1} shows an example of the raw data collected
with an incident energy of 250\,meV and a sample temperature of
10\,K. The data are from a detector bank covering a range of
scattering angle $\phi$ from 3$^\circ$ to 7$^\circ$, i.e.\
$\langle \phi \rangle = 5^\circ$. Spectra from both YbNi$_2$B$_2$C
and Y$_{0.5}$Lu$_{0.5}$Ni$_2$B$_2$C are shown.

To a first approximation, the difference between the scattering
from the two samples is the magnetic scattering, but a straight
subtraction is not an entirely satisfactory way to isolate the
magnetic scattering because (a) the nuclear scattering amplitudes
of Yb, Y and Lu are different, and (b) there are small differences
in the phonon spectra of the two materials, particularly at low
energy. For a more accurate estimate of the non-magnetic
scattering from YbNi$_2$B$_2$C we adopted the following procedure.
First we took the ratio of the energy spectra from YbNi$_2$B$_2$C
and Y$_{0.5}$Lu$_{0.5}$Ni$_2$B$_2$C measured in a high-angle
($\sim$130$^\circ$) detector bank. At high angles the spectrum
measures the phonon density of states weighted by the scattering
power of the elements. The magnetic scattering is negligible
because of the decay of the magnetic form factor with $|{\bf Q}|$.
Multiplication of this high-angle ratio by the low-angle spectrum
of Y$_{0.5}$Lu$_{0.5}$Ni$_2$B$_2$C then gives a good estimate of
the non-magnetic low-angle background of YbNi$_2$B$_2$C. This
procedure is based on the assumption that the non-magnetic
inelastic scattering at low angles is dominated by (elastic +
1-phonon) multiple scattering and therefore closely resembles the
scattering measured at high angles. The validity of this
assumption for high-energy neutrons incident normally on flat
samples whose thickness is small compared to their lateral
dimensions (as is the case here) has been verified previously by
Monte Carlo simulations incorporating realistic model scattering
cross-sections,\cite{Goremychkin-PRB-1993} and also accords with
experience from many other similar experiments. The background
derived this way is included in Fig.\,\ref{fig:1} and is generally
a little higher than the intensity from
Y$_{0.5}$Lu$_{0.5}$Ni$_2$B$_2$C.

\begin{figure}
\includegraphics{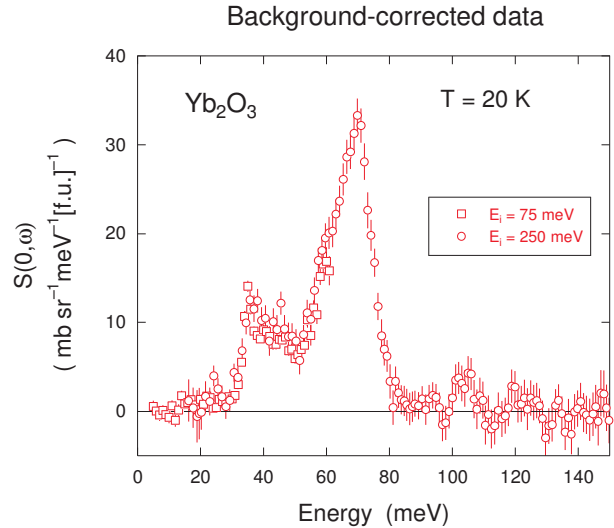}
\caption{Magnetic response function of polycrystalline Yb$_2$O$_3$
measured on the HET time-of-flight spectrometer. The data are
corrected for (i) the non-magnetic background, (ii) the
attenuation of the neutron beam in the sample, and (iii) the
magnetic form factor of Yb$^{3+}$. The normalization is per
formula unit (f.u.) of Yb$_2$O$_3$. Data from runs with two
incident neutron energies, 75\,meV and 250\,meV, are included.
\label{fig:2} }
\end{figure}

As mentioned earlier, the polycrystalline sample of YbNi$_2$B$_2$C
used in these experiments was contaminated with 9\% (by mass) of
Yb$_2$O$_3$ impurity. To correct for the signal from this
Yb$_2$O$_3$ we measured the energy spectra of polycrystalline
samples of pure Yb$_2$O$_3$ and Y$_2$O$_3$, the latter playing the
part of the non-magnetic reference sample. We are not aware of any
previous measurements of the magnetic excitations in Yb$_2$O$_3$,
and so for reference we show in Fig.\,\ref{fig:2} the energy
spectrum of Yb$_2$O$_3$ after correction for the non-magnetic
scattering by the method described above. Corrections have also
been made for the attenuation of the neutron beam in the sample
and for the free-ion magnetic form factor of Yb$^{3+}$. Hence, the
quantity plotted on Fig.\,\ref{fig:2} is $S(0,\omega)$, the
zero-$\bf Q$ magnetic response function. Data from runs with two
different energies has been included. The strongest magnetic
signal in the measured energy range is seen to be centred near
70\,meV.

\begin{figure}
\includegraphics{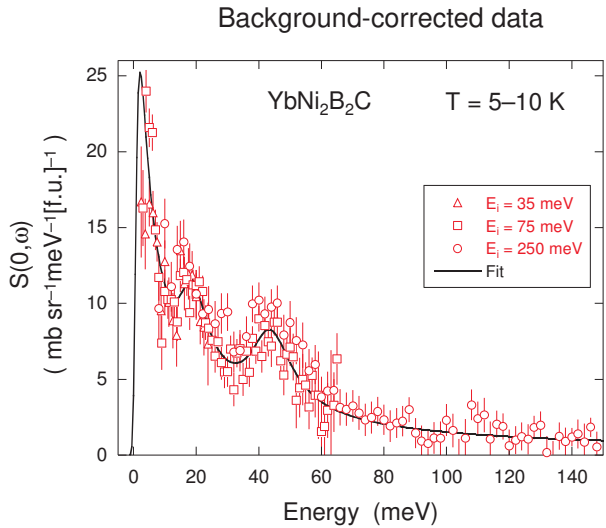}
\caption{Magnetic response function of polycrystalline
YbNi$_2$B$_2$C measured on the HET time-of-flight spectrometer.
The data are corrected for (i) the non-magnetic background, (ii)
the Yb$_2$O$_3$ impurity scattering, (iii) the attenuation of the
neutron beam in the sample, and (iv) the magnetic form factor of
Yb$^{3+}$. The normalization is per formula unit (f.u.) of
YbNi$_2$B$_2$C. Data from runs with three incident neutron
energies, 35\,meV, 75\,meV and 250\,meV, are included. The solid
line is calculated from Eqs.~(\ref{eq:dynamic_res_fn_powder}) and
(\ref{eq:Lorz_spec_fn}) with the parameters given in
Table~\ref{table1}. \label{fig:3} }
\end{figure}

Figure\,\ref{fig:3} shows the zero-$\bf Q$ magnetic response
function of YbNi$_2$B$_2$C derived from the runs with incident
neutron energies 35\,meV, 75\,meV and 250\,meV. To arrive at
Fig.\,\ref{fig:3} we corrected the raw data for (i) the
non-magnetic scattering shown in Fig.\,\ref{fig:1}, (ii) the
Yb$_2$O$_3$ impurity scattering, (iii) the attenuation of the
neutron beam in the sample, and (iv) the magnetic form factor of
Yb$^{3+}$. The good agreement between the results from the three
runs in the energy ranges where they overlap gives us confidence
that the method used to estimate the non-magnetic signal is
reliable. The Yb$_2$O$_3$ impurity correction has little effect
($<$10\%) over most of the energy range, and is only significant
around 70\,meV where it accounts for $\sim$40\% of the raw
magnetic signal in Fig.\,\ref{fig:1}.

\begin{figure}
\includegraphics{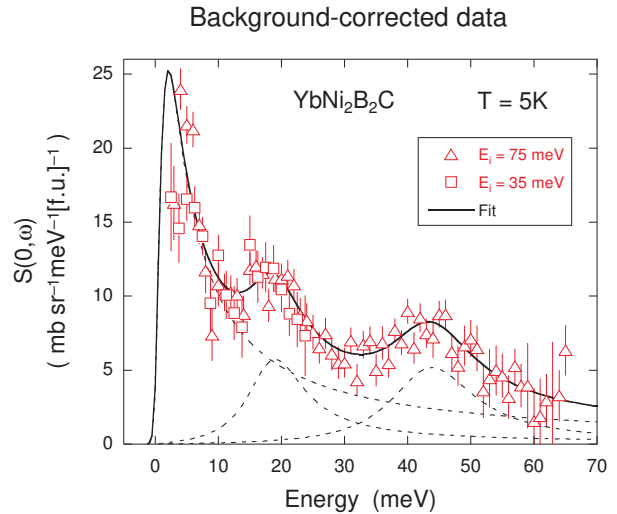}
\caption{Low energy ($\hbar\omega < 70$\,meV) part of
Fig.\,\ref{fig:3} showing the data from runs with incident
energies 35\,meV and 75\,meV. The solid line is calculated from
Eqs.~(\ref{eq:dynamic_res_fn_powder}) and (\ref{eq:Lorz_spec_fn})
with the parameters given in Table~\ref{table1}, and the broken
lines depict the three component peaks that make up the overall
line shape. \label{fig:4} }
\end{figure}

The dynamic magnetic response of YbNi$_2$B$_2$C is seen from
Fig.\,\ref{fig:3} to be broad in energy, extending beyond
150\,meV, but does show some structure. There are three peaks, the
first centred close to zero energy, the second just below 20\,meV,
and the third just above 40\,meV. These peaks can be seen more
clearly in Fig.\,\ref{fig:4}, which highlights the low-energy data
(up to 70\,meV) from the 35\,meV and 75\,meV runs.

To characterize the magnetic response we use
Eq.~(\ref{eq:dynamic_res_fn_powder}) with a phenomenological
spectral-weight function given by
\begin{equation}
F(\omega) = \sum_{i=1}^3 \frac{c_i}{2} \left\{ \frac{\hbar
\Gamma_i / \pi}{(\hbar\omega+E_i)^2+\Gamma_i^2}+ \frac{\hbar
\Gamma_i / \pi}{(\hbar\omega-E_i)^2+\Gamma_i^2}\right\},
\label{eq:Lorz_spec_fn}
\end{equation}
i.e.\ the sum of three pairs of Lorentzian functions centred on
$\pm E_i$ with energy widths (half-width at half maximum)
$\Gamma_i$. The $c_i$ coefficients satisfy $\sum_i c_i = 1$ as
required by the normalization condition,
Eq.~(\ref{eq:spec_wt_fn_norm_ch3}). A Lorentzian function is a
reasonable approximation when the peak broadening is relatively
small and arises from simple relaxational processes. Later on we
will compare the data with the response function calculated by an
approximate method applicable to crystal field transitions in
heavy fermion systems.\cite{ZZF-ZPhysB-1990}

The solid lines drawn on Figs.\,\ref{fig:3} and \ref{fig:4} show
that Eq.~(\ref{eq:Lorz_spec_fn}) can provide a good description of
the data. The parameters used to fit the model response function
to the data are given in Table~\ref{table1}. It is particularly
satisfying that the value $\chi_{\text av} = 8.1\times
10^{-31}$\,m$^3$ per ion (SI units), which corresponds to
0.039\,emu/mol in Gaussian cgs units, matches closely with the
powder susceptibility determined by magnetometry, which at 5\,K is
approximately 0.04\,emu/mol.\cite{Dhar-SSC-1996,Yatskar-PRB-1996}
The justification for the first peak being quasielastic (i.e.\
centred on $\hbar\omega=0$) comes mainly from the low energy data
to be presented shortly. The fit indicates that the 18\,meV and
43\,meV inelastic peaks are broader than the quasielastic peak.
\begin{table}
\caption{\label{table1}Peak parameters of the Lorentzian
spectral-weight function Eq.~(\ref{eq:Lorz_spec_fn}) used to
describe the magnetic response function of YbNi$_2$B$_2$C shown in
Figs.\,\ref{fig:3} and \ref{fig:4}. The value of the
susceptibility determined from the fit is $\chi_{\text av} =
8.1\times 10^{-31}$\,m$^3$ per ion in SI units, which corresponds
to 0.039\,emu/mol in Gaussian cgs units. }
\begin{ruledtabular}
\begin{tabular}{cddd}
\multicolumn{1}{c}{Peak}& \multicolumn{1}{c}{\hspace{30pt}$c_i$\footnotemark[1]} & \multicolumn{1}{r}{$E_i$ \text{(meV)}} & \multicolumn{1}{l}{\hspace{25pt}$\Gamma_i$ \text{(meV)}} \\
\hline\\[-6pt]
1 & 0.91\pm0.03 & 0.0 & 2.1\pm0.1\footnotemark[2]\\
2 & 0.06\pm0.01 & 18.2\pm0.7 & 5.5\pm1.1\\
3 & 0.03\pm0.005 & 43.1\pm0.9 & 8.1\pm1.5\\
\end{tabular}
\end{ruledtabular}
\footnotetext[1]{The error given for $c_i$ is just the statistical
error from the fit, and does not include the $\sim$10\,\%
systematic uncertainty in the absolute calibration of the
data.}\footnotetext[2]{The fitted width of peak 2 is from the
quasielastic fit to the low energy data shown in
Fig.\,\ref{fig:6}.}
\end{table}

\begin{figure}
\includegraphics{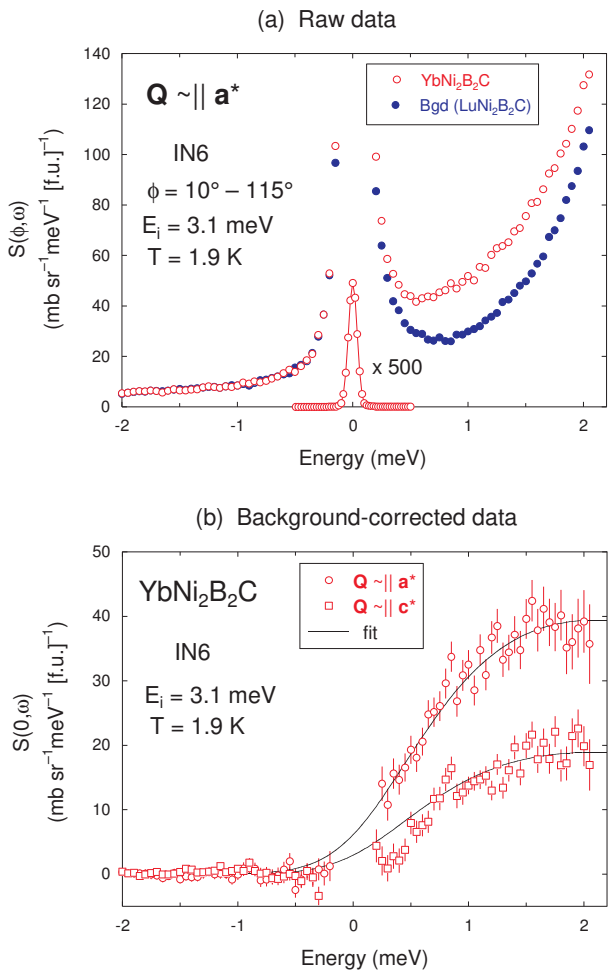}
\caption{(a) Neutron inelastic scattering from single crystal
mosaic samples of YbNi$_2$B$_2$C and LuNi$_2$B$_2$C measured on
the IN6 spectrometer. The elastic peak is shown reduced by a
factor 500 to indicate the energy resolution. (b) Dynamic magnetic
response measured with the scattering vector approximately
parallel to the reciprocal lattice vectors ${\bf a}^{\ast}$ and
${\bf c}^{\ast}$. The data have been corrected for (i) the
non-magnetic background, (ii) the attenuation of the neutron beam
in the sample, and (iii) the magnetic form factor of Yb$^{3+}$.
The solid lines show the model response function of
Eqs.~(\ref{eq:dynamic_res_fn_powder}) and (\ref{eq:Lorz_spec_fn}).
Apart from an overall scale factor the parameters are the same as
those listed in Table~\ref{table1}. \label{fig:5} }
\end{figure}

We now turn to the low energy part of the magnetic response
function. Figure\,\ref{fig:5}(a) displays the neutron inelastic
scattering from the single-crystal samples of YbNi$_2$B$_2$C and
LuNi$_2$B$_2$C measured at a temperature of 1.9\,K on the IN6
spectrometer. In these runs the crystals were oriented so that
across the whole detector bank the scattering vector $\bf Q$ was
approximately parallel to the ${\bf a}^{\ast}$ reciprocal lattice
vector. The two spectra are indistinguishable in the negative
energy region (neutron energy gain scattering), but differ
appreciably at positive energies (neutron energy loss). The
difference in the energy-loss scattering corresponds to the
dynamic magnetic response of the Yb ions in YbNi$_2$B$_2$C. At
negative energies the scattering from the two samples is
essentially the same because the energy-gain magnetic scattering
is strongly suppressed by the factor $\{1+n(\omega)\}$ in the
cross-section,
Eq.~(\ref{eq:dynamic_res_fn_crystal}).

In order to isolate the magnetic scattering we must estimate the
non-magnetic background. On IN6 the sources of background are
different to those described earlier in connection with the HET
measurements. Outside of the energy range $[-1,1]$\,meV the
background is almost sample-independent at low temperatures, while
between $-1$\,meV and 1\,meV there is additional non-magnetic
scattering from the elastic peak. This latter scattering scales
with the amplitude of the elastic peak, which is sample-dependent.
We confirmed these properties of the background by comparing the
LuNi$_2$B$_2$C spectrum with similar measurements from a vanadium
sample and an empty aluminium sample holder. Our procedure to
estimate the non-magnetic background was then as follows. First,
we fitted a polynomial to the parts of the LuNi$_2$B$_2$C spectrum
that lie outside $[-1,1]$\,meV. Next we subtracted this polynomial
from the LuNi$_2$B$_2$C spectrum to separate the elastic peak, and
scaled up this elastic peak to match that in the YbNi$_2$B$_2$C
data. We added the scaled elastic peak back onto the polynomial,
and substituted the result back into the LuNi$_2$B$_2$C spectrum
in place of the original $[-1,1]$\,meV data.

In Fig.\,\ref{fig:5}(b) we show the magnetic response for two
crystal orientations after subtraction of the modified
LuNi$_2$B$_2$C spectrum from the YbNi$_2$B$_2$C spectrum.
Corrections have also been made for the attenuation of the neutron
beam in the sample, and for the magnetic form factor of Yb$^{3+}$.
The magnetic response when $\bf Q$ is approximately parallel to
${\bf a}^{\ast}$ is seen to be a factor two larger than when $\bf
Q$ is approximately parallel to ${\bf c}^{\ast}$. Measurements on
the triple-axis spectrometer IN14, which involves averaging over a
much smaller range of $\bf Q$ than on IN6, were also consistent
with a factor two difference in relative intensities for these
orientations.

\begin{figure}
\includegraphics{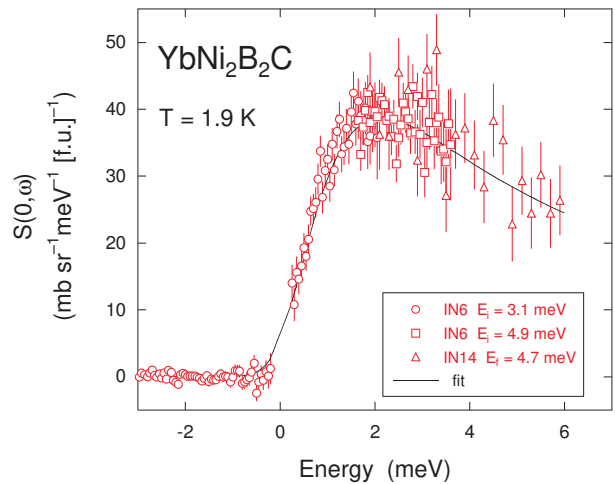}
\caption{Dynamic magnetic response of YbNi$_2$B$_2$C in the low
energy range up to 6\,meV. Data from three different measurements
are shown, each separately corrected for the non-magnetic
background. The lowest-energy data are the ${\bf Q} \sim\parallel
{\bf a}^{\ast}$ points shown on Fig.\,\ref{fig:5}(b), and the
other runs are scaled to match in the energy range where they
overlap. The solid line is the model response function of
Eqs.~(\ref{eq:dynamic_res_fn_powder}) and (\ref{eq:Lorz_spec_fn})
calculated with the parameters listed in Table~\ref{table1}, apart
from $c_1$ which was adjusted to fit the present data.
\label{fig:6} }
\end{figure}

The energy range covered by the data Fig.\,\ref{fig:5}(b) is too
small to allow us to examine the line shape of the low energy
magnetic response in any detail, and so in Fig.\,\ref{fig:6} we
have plotted the ${\bf Q}\sim\parallel{\bf a}^{\ast}$ points from
Fig.\,\ref{fig:5}(b) together with data from two other runs
appended so as to extend the energy range up to 6\,meV. The
additional measurements were made on IN6 with an incident neutron
energy of 4.9\,meV and on IN14 with a fixed final energy of
4.7\,meV. The same single-crystal samples were used for all the
measurements, and after correction for the non-magnetic background
the new data points were scaled so that all three runs matched up
in the energy ranges over which they overlap.

We used the data shown in Fig.\,\ref{fig:6} to refine the model
response function defined by Eqs.~(\ref{eq:dynamic_res_fn_powder})
and (\ref{eq:Lorz_spec_fn}). In this low energy range the 18\,meV
and 43\,meV peaks have negligible weight, as can be seen from
Fig.\,\ref{fig:4}, and the model function is dominated by the
first peak. Allowing the parameters of peak 1 to vary we achieved
the best agreement with a quasielastic peak (i.e.\ $E_1 = 0$) with
width $\Gamma_1 = 2.1\pm0.1$\,meV. Attempts to force the peak to
be inelastic ($E_1 > 0$) resulted in progressively poorer fits as
the centre of the Lorentzian was shifted to higher energies. These
values of $E_1$ and $\Gamma_1$ were then used as fixed parameters
in a further fit to the polycrystalline data (Figs.\,\ref{fig:3}
and \ref{fig:4}) which finally established the other model
parameters listed in Table~\ref{table1}. We note that a powder
average of the single-crystal data shown in Fig.\,\ref{fig:5}(b)
agrees to within 25\% of the the model response function displayed
on Figs.\,\ref{fig:3} and \ref{fig:4}, which is an acceptable
margin of error given the uncertainties in latter fit and in the
absolute calibration of the data. This agreement gives us further
confidence in the $c_i$ and $\chi_{\rm av}$ amplitude parameters.

We mention finally that on the IN14 triple-axis spectrometer we
examined the low-energy scattering in a series of energy scans at
fixed ${\bf Q}$, with ${\bf Q}$ chosen at different points along
the main symmetry directions $(1,0,0)$, $(1,1,0)$ and $(0,0,1)$,
as well as at some off-symmetry positions. As already mentioned,
the intensity of the scattering was found to vary with the
direction of ${\bf Q}$ (smallest when ${\bf Q}$ is parallel to the
${\bf c}^{\ast}$ direction and constant within the
$a^{\ast}b^{\ast}$ plane), but to within the statistical precision
of the data we did not observe any change in the line shape with
${\bf Q}$. The measurements allow us to place an upper limit of
0.5\,meV on the extent of any dispersion in the low energy
magnetic response.

\section{\label{sec:inel_disc}Analysis and discussion}

These measurements have revealed that the magnetic excitation
spectrum of YbNi$_2$B$_2$C contains three peaks, centred
approximately on 0\,meV, 18\,meV and 43\,meV, with widths of
several meV. The broadening is such that there is significant
overlap of the peaks, but not enough to prevent the peaks from
being resolved.

What does this tell us about the Yb electronic state in
YbNi$_2$B$_2$C? In general, the properties of Yb intermetallic
compounds are influenced by a partial hybridization of the Yb 4$f$
electrons with conduction or valence electron states of the
host.\cite{LoewenhauptFischer-1993} When the hybridization is very
weak the Yb ions are trivalent with localized 4$f$ electrons
carrying a magnetic moment. The dominant external interaction is
then with the crystalline electric field (CEF), which splits the
$^2$F$_{7/2}$ ground state term of Yb$^{3+}$ into a manifold of
levels. Transitions between these CEF levels give rise to sharp
peaks in the magnetic excitation spectrum. The opposite extreme is
when the hybridization is much stronger than the CEF. This results
in valence fluctuations of the Yb ions, and the corresponding
dynamic magnetic response is very broad in energy.

The case of YbNi$_2$B$_2$C is seemingly part-way between these
extremes. However, the fact that we observe inelastic peaks rather
than a single broad response suggests that a description in terms
of localized 4$f$ electrons in a CEF is the better starting point.
This scenario is consistent with the X-ray absorption spectrum
measured at the L$_{\rm III}$ edge, which indicated a stable 3+
ionization state for the Yb ions.\cite{Dhar-SSC-1996}

Given a stable Yb valence, the broadening of the CEF transitions
is expected to arise from the same strong spin fluctuations that
dominate the electronic heat capacity at low temperatures. To see
if this is the case we consider the relation
\begin{equation}
\gamma = \frac{\nu\pi k_{\rm
B}^2}{3\Gamma},\label{eq:Edwards-Lonzarich}
\end{equation}
obtained by Edwards and
Lonzarich\cite{EdwardsLonzarich-PhilMag-1992} for the
low-temperature electronic heat capacity $\gamma T$ of a system of
fluctuating spins described by a set of over-damped harmonic
oscillators with a $\bf Q$-independent relaxation width $\Gamma$.
Here, $\nu$ is the number of fluctuating degrees of freedom.
Eq.~(\ref{eq:Edwards-Lonzarich}) has been shown to hold for a
range of materials in which strong spin fluctuations are known to
be important.\cite{Hayden-PRL-2000} For an effective
spin--$\frac{1}{2}$ system, such as we have here, the spin state
is defined by one component and so $\nu = 1$. If we take $\Gamma
 = 2.1$\,meV, the quasielastic width, then
Eq.~(\ref{eq:Edwards-Lonzarich}) predicts $\gamma =
0.36$\,J\,mol$^{-1}$\,K$^{-2}$, which is roughly 30\% smaller than
found experimentally for YbNi$_2$B$_2$C. This reasonable level of
agreement supports the hypothesis that spin fluctuations account
for most of the low temperature electronic heat capacity in
YbNi$_2$B$_2$C. Exchange interactions between the Yb ions can be
ruled out as a significant contributor to the CEF broadening
because the magnetic excitations do not exhibit any observable
dispersion. Furthermore, if Yb--Yb interactions were important
then dilution of the Yb ions by non-magnetic Lu should make the
magnetic excitations sharper and therefore easier to observe, but
neutron measurements on Yb$_{0.1}$Lu$_{0.9}$Ni$_2$B$_2$C have
failed to detect any such sharpening.\cite{Gasser-PhysicaB-1997}

We will now examine the CEF in YbNi$_2$B$_2$C more quantitatively.
CEF energy spectra for $R$Ni$_2$B$_2$C ($R$ = Dy, Ho, Er and Tm)
measured by neutron spectroscopy have been presented and analyzed
in detail by Gasser {\it et al}.\cite{Gasser-PhysicaB-1997,
Gasser-ZPhysB-1996} For the 4/$mmm$ (D$_{4h}$) point symmetry of
the $R$ site the angular dependence of the CEF is described by
five parameters (essentially coefficients of spherical harmonics).
Gasser {\it et al}. derived values for the CEF parameters for $R$
= Dy, Ho, Er and Tm by comparing quantities calculated from the
CEF model with the available spectroscopic and magnetic data.
After correction for the expected dependence on the ion size there
is little variation in the CEF parameters for each ion. Hence,
Gasser {\it et al}. extrapolated the CEF parameters to other $R$
ions and predicted the CEF splittings.

For the case $R$ = Yb, the extrapolations predict that the
$^2$F$_{7/2}$ term splits into four Kramers' doublets. The
predicted energies and wavefunctions of the doublets are given in
Table~\ref{table2}. The neutron scattering cross-sections for
transitions (i) within the ground state, and (ii) from the ground
state to each of the three excited doublets are calculated to be
(i) 497\,mb\,sr$^{-1}$\,[f.u.]$^{-1}$, and (ii) 260, 228 and
9\,mb\,sr$^{-1}$\,[f.u.]$^{-1}$. From Fig. \ref{fig:4} it is
evident on the one hand that the observed CEF splitting is nearly
twice that predicted by the extrapolated CEF model, but on the
other that the relative spacings of energy levels and their
transition intensities match quite closely with the predictions.
We see also that the absence of a third excited CEF peak in the
experimental spectrum is explained by its small spectral weight
and by its close proximity to the second excited CEF peak.
\begin{table}
\caption{\label{table2}Energy levels, symmetries and wavefunctions
of the CEF doublets in YbNi$_2$B$_2$C calculated from the CEF
extrapolated from TmNi$_2$B$_2$C.\cite{Gasser-ZPhysB-1996} The CEF
parameters (Stevens' operator notation) are $B_2^0 =
-3.95\times10^{-1}$, $B_4^0 = -3.46\times10^{-3}$, $B_4^4 =
1.01\times10^{-1}$, $B_6^0 = -1.77\times10^{-4}$, $B_6^4 =
2.09\times10^{-3}$\,meV.}
\begin{ruledtabular}
\begin{tabular}{rcc}
Energy& Symmetry & Wavefunction\footnotemark[2] \\ (meV) &
label\footnotemark[1] &
$\sum_{m_J}a_{m_J}|m_J\rangle$\\[4pt]
\hline \\[-6pt] 0.0 & $\Gamma_6^-$ & $0.902|\pm
7/2\rangle -0.431|\mp 1/2\rangle$ \\
7.9 & $\Gamma_7^-$ & $0.717|\mp 3/2\rangle - 0.697|\pm
5/2\rangle$ \\
24.3 & $\Gamma_6^-$ & $0.431|\pm
7/2\rangle + 0.902|\mp 1/2\rangle$ \\
26.0 & $\Gamma_7^-$ & $0.697|\mp 3/2\rangle + 0.717|\pm
5/2\rangle$ \\
\end{tabular}
\end{ruledtabular}
\footnotetext[1]{Irreducible representations of the 4/$mmm$
(D$_{4h}$) double group, as defined in Ref.\
\onlinecite{Koster}.}\footnotetext[2]{ $|m_J\rangle$ is shorthand
for $|J,m_J\rangle$, where $J=7/2$ is the combined spin and
orbital angular momentum quantum number of the $^2$F$_{7/2}$ term,
and $m_J$ is the magnetic quantum number. We assume negligible
admixture of states from the $^2$F$_{5/2}$ term. In the
$R$Ni$_2$B$_2$C structure the CEF quantization axis is parallel to
the crystallographic $c$ axis.}
\end{table}

Encouraged, we now consider the anisotropy in the magnetic
scattering (Fig.\ \ref{fig:5}b) and bulk magnetic susceptibility.
The neutron scattering cross-section for a transition between two
CEF levels $i$ and $j$ depends on $|\langle
i|\hat{J}_{\alpha}|j\rangle |^2$, the squared matrix elements of
the angular momentum operator $\hat{J}_{\alpha}$ ($\alpha = x, y,
z$).\cite{Lovesey} The $(1-\hat{Q}_{\alpha}^2)$ factor in Eq.\
\ref{eq:dynamic_res_fn_crystal} means that when ${\bf Q}
\parallel {\bf c}^{\ast}$ the intensity is proportional to $|\langle
i|\hat{J}_x|j\rangle |^2+|\langle i|\hat{J}_y|j\rangle |^2$, and
when ${\bf Q} \parallel {\bf a}^{\ast}$ the intensity is
proportional to $|\langle i|\hat{J}_y|j\rangle |^2 + |\langle
i|\hat{J}_z|j\rangle |^2$. In tetragonal symmetry $|\langle
i|\hat{J}_x|j\rangle |^2 = |\langle i|\hat{J}_y|j\rangle |^2$, and
assuming the low-energy scattering comes mainly from scattering
within the ground-state doublet we deduce from the relative
intensities in Fig.\ \ref{fig:5}b that $|\langle
0|\hat{J}_z|0\rangle |^2 \approx 3|\langle 0|\hat{J}_x|0\rangle
|^2$. However, from the wavefunctions of the extrapolated CEF
model (Table~\ref{table2}) we find $|\langle 0|\hat{J}_z|0\rangle
|^2 \approx 50|\langle 0|\hat{J}_x|0\rangle |^2$. In other words,
the sense of the anisotropy predicted by the extrapolation is
correct, but the observed degree of anisotropy is much smaller
than predicted by this CEF-only model.

\begin{figure}
\includegraphics{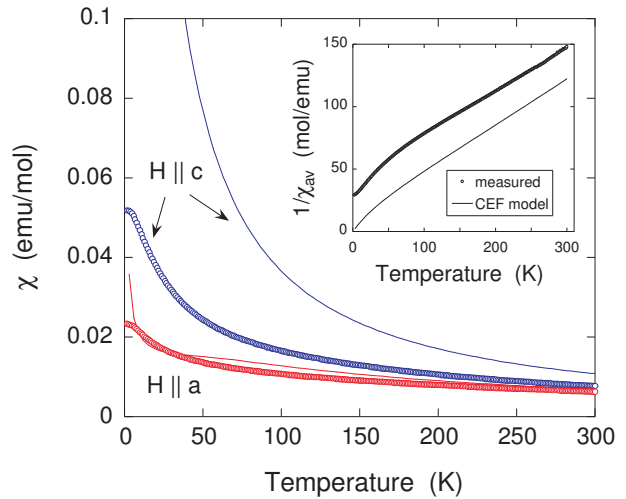}
\caption{A comparison of the measured magnetic susceptibility of
YbNi$_2$B$_2$C (open circles) with the susceptibility calculated
from the scaled CEF model (lines). The experimental data has been
taken from Avila {\it at al.}\cite{Avila-cond-mat-2002} The inset
shows the inverse of the powder-averaged susceptibility.
\label{fig:7} }
\end{figure}

The same conclusion can be drawn from the bulk susceptibility.
Figure\ \ref{fig:7} compares the susceptibility calculated from
the extrapolated CEF model with the single crystal data reported
in Ref.\ \onlinecite{Avila-cond-mat-2002}. Again, the sense of
anisotropy is in accord, but the model predicts a much greater
degree of anisotropy than observed experimentally. A further
observation is that the measured powder-averaged susceptibility
$\chi_{\text av}$ is less than the calculated $\chi_{\text av}$,
most of the difference being in the $H \parallel c$ component.
Attempts to refine the CEF model indicated that a reduction in
magnetic anisotropy could be achieved only at the expense of a
poorer agreement with the scattering intensities and without
significant change in $\chi_{\text av}$. Hence, we were not able
to find an acceptable alternative model that was substantially
different from the extrapolated CEF model given in
Table~\ref{table2}.

The discrepancy in $\chi_{\text av}$ increases dramatically at low
temperatures, but when compared on a plot of $\chi_{\text
av}^{-1}$ vs $T$, as done in the inset to Fig.\ \ref{fig:7}, it
becomes apparent that over much of the temperature range the
measured and predicted curves have the same slope. This suggests
the influence of an effective antiferromagnetic exchange
interaction on the Yb ions, a likely source of which is the
on-site Kondo interaction (we have already ruled out the existence
of measurable inter-site antiferromagnetic interactions). In the
following analysis, therefore, we consider the addition of a Kondo
interaction to the CEF potential established above, and use
approximate methods to see whether this model leads to a better
description of the experimental results.

\begin{figure}
\includegraphics{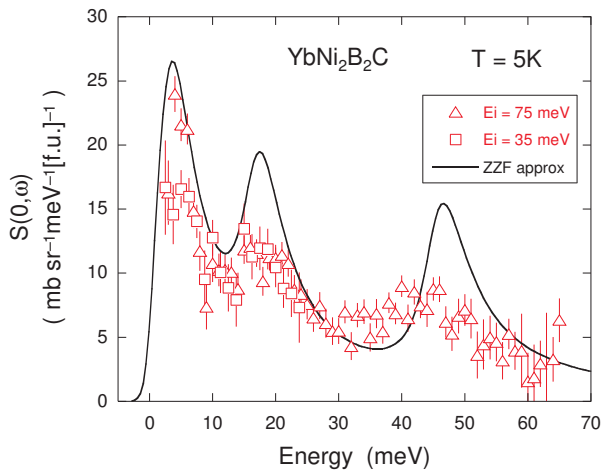}
\caption{Dynamic magnetic response of YbNi$_2$B$_2$C calculated by
the method of Ref. \onlinecite{ZZF-ZPhysB-1990} compared with the
experimental data from Fig.\ \ref{fig:4}. The model contains the
CEF parameters extrapolated from TmNi$_2$B$_2$C scaled by a factor
of 1.8, and a Kondo temperature $T_{\rm K}=25$\,K. \label{fig:8} }
\end{figure}

We consider first the influence of a Kondo interaction on the
dynamic magnetic response. Previously, a satisfactory description
of the measured neutron magnetic scattering spectra of several Yb
Kondo lattice compounds has been obtained from a simple
approximate solution to the Anderson impurity
Hamiltonian.\cite{ZZF-ZPhysB-1990,Polatsek-ZPhysB-1992} Within
this scheme (the ZZF approximation\cite{ZZF-ZPhysB-1990}) one can
calculate the single-ion dynamic magnetic response using as input
the CEF parameters, the Kondo temperature $T_{\rm K}$, and the
number of $f$ electron holes $n_f$. To apply this method to
YbNi$_2$B$_2$C we scaled the CEF parameters extrapolated from
TmNi$_2$B$_2$C by a factor of 1.8 to match the energy scale of the
observed CEF splitting, and set $n_f=0.95$, close to unity to
reflect that the Yb valence is close to 3+. When $0.9\leq n_f \leq
1$ the results are not sensitive to the precise value chosen for
$n_f$. We fixed $T_{\rm K}=25$\,K from the relation $k_{\rm
B}T_{\rm K}=\Gamma = 2.1$\,meV from the quasielastic
width.\cite{comment_on_TK} Fig.\ \ref{fig:8} shows the response
calculated by the ZZF approximation together with the HET data
from Fig.\ \ref{fig:4}. Given that no attempt has been made to
refine the CEF model the level of agreement as far as the peak
positions and relative intensities are concerned is surprisingly
good. This indicates that the scaled CEF model is a good starting
point for a description of the CEF in YbNi$_2$B$_2$C. What is also
clear from Fig.\ \ref{fig:8} is that the ZZF approximation
appreciably underestimates the widths of the 18\,meV and 43\,meV
peaks. A complete understanding of the lineshape is evidently
lacking\cite{footnote2}

We next turn to the powder-averaged magnetic susceptibility, which
is connected to the dynamic magnetic response by Eq.\
(\ref{eq:dynamic_res_fn_powder}). Integrating the ZZF dynamic
response given in Fig.\ \ref{fig:8} we obtain $\chi_{\rm
av}=0.036$\,emu/mol (or $1/\chi_{\rm av}=28$\,mol/emu) at $T =
5$\,K. This result is very close to the experimental value, as may
be seen from the inset to Fig.\ \ref{fig:7}. At higher
temperatures, however, we find that the ZZF model systematically
exceeds the experimental susceptibility as temperature increases.
In order to obtain agreement it is necessary to increase $T_{\rm
K}$ in the ZZF model systematically with temperature. We find that
the $T_{\rm K}$ needed in the ZZF model to achieve agreement with
experiment increases smoothly with temperature, eventually
saturating at $T_{\rm K}=130$\,K at high temperatures.

Such a renormalization of the Kondo interaction with temperature
is well known in systems where the CEF splitting is larger than
the low temperature $T_{\rm K}$, and has already been discussed
for YbNi$_2$B$_2$C by Rams {\it et al}.\cite{Rams-JMMM-2000} in
connection with measurements of the electric field gradient. The
effective Kondo interaction renormalizes down with decreasing
temperature as a result of the reduction in the effective
degeneracy of the Yb$^{3+}$ ion due to the CEF splitting.
According to theory,\cite{Hanzawa-JMMM-1985} the effective Kondo
temperature at high temperature $T_{\rm K}^h$ is related to that
at low temperature $T_{\rm K}^0$ by
\begin{equation}
T_{\rm K}^h \approx (\Delta_1\Delta_2\Delta_3T_{\rm
K}^0)^{1/4},\label{eq:TK}
\end{equation}
where the $\Delta_i$ are the energies (in Kelvin) of the excited
CEF levels relative to the ground state. Eq.~(\ref{eq:TK}) is
valid if $T_{\rm K}^0 \ll \Delta_i$, a condition that is satisfied
here. If we set $T_{\rm K}^0 = 25$\,K and use the observed CEF
levels for $\Delta_i$ then Eq.\ (\ref{eq:TK}) predicts $T_{\rm
K}^h\approx 190$\,K, which is comparable to the value $T_{\rm
K}=130$\,K required in the ZZF model to match the observed high
temperature susceptibility.

Finally, we comment on the neutron scattering measurements
reported earlier by Sierks {\it et al}.\cite{Sierks-PhysicaB-1999}
In Ref. \onlinecite{Sierks-PhysicaB-1999} the authors investigated
both the low and high energy part of the excitation spectrum using
a polycrystalline sample of YbNi$_2$B$_2$C. Without a non-magnetic
reference spectrum they were not able to find the broad inelastic
magnetic peaks centred on 18\,meV and 43\,meV we have identified
here. They did, however, observe two inelastic magnetic features
at low energies. One of these is a broad shoulder on the side of
the elastic peak extending up to approximately 1\,meV, which
developed into a quasielastic peak with increasing temperature,
and the other a broader peak centred near 3.5\,meV. We believe the
latter feature is what we have identified here as quasielastic
scattering within the ground state doublet, but concerning the
scattering below 1\,meV there is no sign of this in our data, as
is clear from Fig.\ \ref{fig:5}b. The possibility that this
scattering is absent from our data because it is highly localized
in reciprocal space can be discounted because the large detector
bank of the IN6 spectrometer used to collect the single-crystal
data meant that the two scans shown in Fig.\ \ref{fig:5}b include
data from all parts of the Brillouin zone.

\begin{figure}
\includegraphics{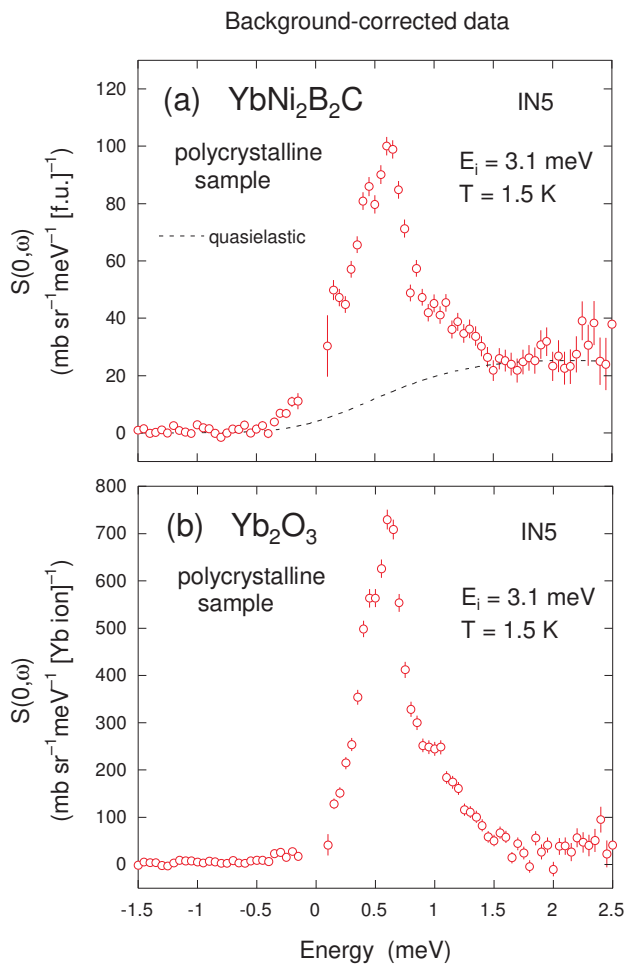}
\caption{Low-energy magnetic excitations measured by neutron
inelastic scattering from polycrystalline samples of (a)
YbNi$_2$B$_2$C and (b) Yb$_2$O$_3$. The data have been corrected
for (i) the non-magnetic background, (ii) the attenuation of the
neutron beam, and (iii) the magnetic form factor of Yb$^{3+}$. The
data show that the peak centred near 0.5\,meV in the
YbNi$_2$B$_2$C data originates from the $\sim$9\% Yb$_2$O$_3$
impurity present in the sample.\label{fig:9} }
\end{figure}

The most likely explanation, therefore, is that the magnetic
scattering observed by Sierks {\it et al}.\ below 1\,meV
originates from an impurity in their sample, probably Yb$_2$O$_3$.
To support this argument we show in Fig.\ \ref{fig:9}a the
background-corrected spectrum of our own polycrystalline sample of
YbNi$_2$B$_2$C and in Fig.\ \ref{fig:9}b the background-corrected
spectrum of polycrystalline Yb$_2$O$_3$, both spectra measured on
the IN5 spectrometer under the same experimental conditions to
those used by Sierks {\it et al}. The peak centred at 0.5\,meV
present in both spectra is clearly a signature of
Yb$_2$O$_3$,\cite{Yb2O3} and its intensity ratio in Figs.\
\ref{fig:9}a and b (when scaled by the respective formula masses)
is consistent with the 9\% mass fraction of Yb$_2$O$_3$ found in
the phase analysis of our YbNi$_2$B$_2$C sample. Assuming the same
impurity was responsible for the 0.5\,meV signal in Sierks {\it et
al}.'s data we estimate that their YbNi$_2$B$_2$C sample contained
$\sim$4\% Yb$_2$O$_3$.

\section{\label{sec:conc}Conclusions}

The results and analysis presented here reveal that the magnetic
properties of YbNi$_2$B$_2$C are principally determined by CEF and
Kondo interactions acting on the Yb$^{3+}$ ions, with the CEF
about an order of magnitude larger than the Kondo energy scale
($T_{\rm K}\approx 10 $\,K from the single-impurity expressions
for the heat capacity and susceptibility, and $T_{\rm K}\approx 25
$\,K from the quasielastic width)\cite{comment_on_TK}. Any
magnetic coupling between Yb ions is negligible. The CEF potential
is roughly a factor 2 greater than found in other $R$Ni$_2$B$_2$C
compounds, presumably amplified by hybridization of the Yb 4$f$
electrons with neighbouring atomic orbitals. Being an
electrostatic interaction, hybridization is expected to modify the
CEF to some degree; here, the main effect seems to be a uniform
enhancement of the CEF without significant change in its
anisotropy.

The ZZF approximation scheme for the Anderson impurity Hamiltonian
succeeds in connecting the low temperature magnetic susceptibility
and dynamic magnetic response, and thus provides a microscopic
basis on which to interpret the properties of YbNi$_2$B$_2$C. The
effective Kondo scale increases with temperature due to the
thermal population of excited CEF states and associated increase
in orbital degeneracy and effective moment. This mechanism
explains why the Weiss temperature $\theta_{\rm p}$ derived from
the high temperature susceptibility is much larger than the value
of the Kondo temperature deduced from various low-temperature
properties. In effect, $\theta_{\rm p}$ is a measure of the high
temperature Kondo scale. It would be interesting to perform
inelastic neutron scattering measurements at higher temperatures
to test the ZZF prediction of a marked broadening of the dynamic
magnetic response with increasing $T_{\rm K}$.

Our work has shown that in most respects YbNi$_2$B$_2$C can be
understood in terms of conventional ideas for Kondo-lattice
intermetallic compounds. However, in one respect, not mentioned so
far, there are indications that less conventional physics may be
needed. This concerns the low temperature thermodynamic, transport
and magnetic data which, if one examines the literature closely,
are seen to exhibit significant deviations from Fermi liquid
behaviour. Specifically, the susceptibility does not saturate
below $T_{\rm K}$,\cite{Yatskar-PRB-1996,Avila-cond-mat-2002} the
sommerfeld coefficient $\gamma$ continues to increase as $T$ tends
to zero,\cite{Dhar-SSC-1996,Yatskar-PRB-1996} and the low
temperature resistivity could be consistent with a $T^n$
dependence with $n<2$.\cite{Yatskar-PRB-1996}

These temperature dependences in fact bear a strong resemblence to
non-Fermi-liquid effects exhibited by YbRh$_2$Si$_2$, a material
recently proposed to be the first example of a clean Yb compound
showing quantum-critical behaviour above a low-lying
antiferromagnetic transition.\cite{Trovarelli-PRL-2000} We
conjecture, therefore, that at zero temperature YbNi$_2$B$_2$C may
be close to a quantum-critical point \emph{on the non-magnetic
side}. The ground state would then contain strong quantum
fluctuations of the spins, and deviations from the Kondo model
should be expected. Indeed, the fact that
Eq.~(\ref{eq:Edwards-Lonzarich}), which is based on a
phenomenological model for the spin fluctuation spectrum, succeeds
in connecting the experimental values of $\gamma$ and $\Gamma$ (to
within 30\%), whereas the single-impurity Kondo model does not
(the $T_{\rm K}$ values deduced from $\gamma$ and $\Gamma$ differ
by a factor of 2.5) may be evidence for a breakdown of the
conventional heavy fermion scenario. Unfortunately, the present
experiments are not sensitive enough to shed any further light on
this possibility, but the indications are clear enough to suggest
that future investigations into the low temperature properties of
YbNi$_2$B$_2$C may prove to be fruitful.

\begin{acknowledgments}
We thank Clemens Ritter for help with the multiphase structure
refinement of the polycrystalline YbNi$_2$B$_2$C sample, and
Martin Lees, Alistair Campbell and Don Paul for help with the
preparation of the polycrystalline samples at the University of
Warwick. We are grateful to Marcos Avila for supplying the
susceptibility data shown in Fig.\ \ref{fig:7} prior to
publication of Ref.\ \onlinecite{Avila-cond-mat-2002}, and to
Michael Loewenhaupt for a critical reading of the manuscript.
Financial support was provided by the Engineering and Physical
Sciences Research Council of Great Britain. Ames Laboratory is
operated for the U.S. Department of Energy by Iowa State
University under Contract No. W-7405-Eng-82. This work was
supported by the Director for Energy Research, Office of Basic
Energy Sciences.
\end{acknowledgments}


\begin{references}

\bibitem{Nagarajan-PRL-1994}
R. Nagarajan, C. Mazumdar, Z. Hossain, S.K. Dhar, K.V.
Gopalakrishnan, L.C. Gupta, C. Godart, B.D. Padalia, and R.
Vijayaraghavan, Phys. Rev. Lett. {\bf 72}, 274 (1994).

\bibitem{Cava-Nature-1994}
R.J. Cava, H. Takagi, H.W. Zandbergen, J.J. Krajewski, W.F. Peck,
Jr., R.B. van Dover, R.J. Felder, T. Siegrist, K. Mizuhahi, J.O.
Lee, H. Eisaki, S.A. Carter, and S. Uchida, Nature (London) {\bf
367}, 252 (1994).

\bibitem{Muller-RPP-2001}
K-H. M\"{u}ller and V.N. Narozhnyi, Rep. Prog. Phys. {\bf 64}, 943
(2001).

\bibitem{Canfield-PhysicsTD-1998}
P.C. Canfield, P.L. Gammel, and D.J. Bishop, Physics Today,
October 1998, 40 (1998).

\bibitem{Lai-PRB-1995}
C.C. Lai, M.S. Lin, Y.B. You, H.C. Ku, Phys. Rev. B {\bf 51}, 420
(1995).

\bibitem{Massalami-PhysicaC-1998}
M. El Massalami, R.E. Rapp, and G.J. Nieuwenhuys, Physica C {\bf
304}, 184 (1998).

\bibitem{Dhar-SSC-1996}
S.K. Dhar, R. Nagarajan, Z. Hossain, E. Tominez, C. Godart, L.C.
Gupta, and R. Vijayaraghavan, Solid State Commun. {\bf 98}, 985
(1996).

\bibitem{Yatskar-PRB-1996}
A. Yatskar, N.K. Budraa, W.P. Beyermann, P.C. Canfield, and S.L.
Bud'ko, Phys. Rev. B {\bf 54}, R3772 (1996).

\bibitem{Avila-cond-mat-2002}
M.A. Avila, S.L. Bud'ko, and P.C. Canfield, Phys. Rev. B {\bf 66},
132504 (2002).

\bibitem{Sala-PRB-1997}
R. Sala, F. Borsa, E. Lee, and P.C. Canfield, Phys. Rev. B {\bf
56}, 6195 (1997).

\bibitem{Bonville-EPJ-1999}
P. Bonville, J.A. Hodges, Z. Hossain, R. Nagarajan, S.K. Dhar,
L.C. Gupta, E. Alleno, and C. Godart, Eur. Phys. J B {\bf 11}, 377
(1999).


\bibitem{Boothroyd-NATOARW-2001}
A.T. Boothroyd, J.P. Barratt, S.J.S. Lister, A.R. Wildes, P.C.
Canfield, and R.I. Bewley, in {\it Rare Earth Transition Metal
Borocarbides (Nitrides): Superconductivity, Magnetic and Normal
State Properties}, edited by K.-H. M\"{u}ller and V. Narozhnyi
(Kluwer, Netherlands, 2001), p. 163.

\bibitem{Sierks-PhysicaB-1999}
C. Sierks, M. Loewenhaupt, P. Tils, J. Freudenberger, K.-H.
M\"{u}ller, C.-K. Loong, and H. Schober, Physica B {\bf 259--261},
592.

\bibitem{Xu-PhysicaB-1994}
Ming Xu, P.C. Canfield, J.E. Ostenson, D.K. Finnemore, B.K. Cho,
Z.R. Wang, and D.C. Johnston, Physica C {\bf 227}, 321 (1994).

\bibitem{Cho-PRB-1995}
B.K. Cho, P.C. Canfield, L.L. Miller, D.C. Johnston, W.P.
Beyermann, and A. Yatskar, Phys. Rev. B {\bf 52}, 3684 (1995).


\bibitem{Lovesey}
S.W. Lovesey, {\it Theory of neutron scattering from condensed
matter} (Oxford University Press, Oxford, 1986).

\bibitem{footnote1}
Here we use S.I. units, and define the single-ion susceptibility
by ${\bf m} = \chi{\bf H}$, where ${\bf m}$ is the magnetic moment
(units of A\,m$^2$ or J\,T$^{-1}$) induced on an ion by an applied
field ${\bf H}$ (units of A\,m$^{-1}$).

\bibitem{Goremychkin-PRB-1993}
E.A. Goremychkin and R. Osborn, Phys. Rev. B {\bf 47}, 14280
(1993).


\bibitem{ZZF-ZPhysB-1990}
G. Zwicknagl, V. Zevin, and P. Fulde, Z. Phys. B {\bf 79}, 365
(1990).

\bibitem{LoewenhauptFischer-1993}
M. Loewenhaupt and K.H. Fischer, in {\it Handbook of Magnetic
Materials}, edited by K.H.J. Buschow (Elsevier, Amsterdam, 2001),
Vol. 7, p. 503.

\bibitem{EdwardsLonzarich-PhilMag-1992}
D.M. Edwards and G.G. Lonzarich, Phil. Mag. B {\bf 65}, 1185
(1992).

\bibitem{Hayden-PRL-2000}
S.M. Hayden, R. Doubble, G. Aeppli, T.G. Perring, and E. Fawcett,
Phys. Rev. Lett. {\bf 84}, 999 (2000).

\bibitem{Gasser-PhysicaB-1997}
U. Gasser, P. Allenspach, and A. Furrer, Physica B {\bf 241--243},
789 (1997). In this work the magnetic scattering was too weak to
separate from the background, but had the magnetic peaks become
sharper with dilution then they most likely would have been large
enough to observe above background.

\bibitem{Gasser-ZPhysB-1996}
U. Gasser, P. Allenspach, F. Fauth, W. Henggeler, J. Mesot, A.
Furrer, S. Rosenkranz, P. Vorderwisch, and M. Buchgeister, Z.
Phys. B {\bf 101}, 345 (1996).

\bibitem{Koster}
G.F. Koster, J.O. Dimmock, R.G. Wheeler, and H. Statz, {\it
Properties of the Thirty-Two Point Groups} (Cambridge, MA: MIT
Press, 1963).

\bibitem{Polatsek-ZPhysB-1992}
G. Polatsek and P. Bonville, Z. Phys. B {\bf 88}, 189 (1992).

\bibitem{comment_on_TK}
We note that the value $T_{\rm K}\approx 25$\,K derived from the
quasielastic width is more than double the established value
($T_{\rm K}\approx 10 $\,K) obtained via the single-impurity Kondo
model from low temperature measurements of the heat capacity and
susceptibility.\cite{Yatskar-PRB-1996} Because YbNi$_2$B$_2$C is a
concentrated system it may not be reasonable to expect these two
values to agree, but the discrepancy deserves further
investigation.

\bibitem{footnote2}
According to a number of theoretical approaches the low-energy
magnetic response, which is a quasielastic Lorentzian at high
temperatures, is predicted to become inelastic below a temperature
comparable to $T_{\rm K}$ due the formation of the Kondo singlet
ground state.\cite{ZZF-ZPhysB-1990, Kondolineshapes} All peaks in
the dynamic magnetic response are then inelastic, and represent
transitions from the Kondo singlet to CEF-like excited states.
Evidence for a crossover from a quasielastic to an inelastic line
shape has been found in the low-energy spectrum of
YbAgCu$_4$,\cite{Severing-PRB-1990} but the present low-energy
data (Fig.\ \ref{fig:6}) are of insufficient precision to search
for the predicted small deviations in line shape.

\bibitem{Kondolineshapes}
P. Schlottmann, Phys. Rev. B {\bf 25}, 2371 (1982); Y. Kuramoto
and E. M\"{u}ller-Hartmann, J. Magn. Magn. Mater. {\bf 52}, 122
(1985); N.E. Bickers, D.L. Cox, and J.W. Wilkins, Phys. Rev. B
{\bf 36}, 2036 (1987).

\bibitem{Severing-PRB-1990}
A. Severing, A.P. Murani, J.D. Thompson, Z. Fisk, and C.-K. Loong,
Phys. Rev. B {\bf 41}, 1739 (1990).

\bibitem{Yb2O3}
At temperatures above 1.5\,K the 0.5\,meV peak was found to
develop into a quasielastic line shape, just as described by
Sierks {\it et al}. Yb$_2$O$_3$ orders antiferromagnetically below
$T_{\rm N}=2.3$\,K [see R.M. Moon, W.C. Koehler, H.R. Child, and
L.J. Raubenheimer, Phys. Rev. {\bf 176}, 722 (1968)], and so this
peak most likely corresponds to the transition between the two
components of the Yb ground state doublet split by the static
exchange field present below $T_{\rm N}$.

\bibitem{Rams-JMMM-2000}
M. Rams, K. Kr\'{o}as, P. Bonville, J.A. Hodges, Z. Hossain, R.
Nagarajan, S.K. Dhar, L.C. Gupta, E. Alleno, and C. Godart, J.
Magn. Magn. Mater. {\bf 219}, 15 (2000).

\bibitem{Hanzawa-JMMM-1985}
K. Hanzawa, K. Yamada, and K. Yoshida, J. Magn. Magn. Mater. {\bf
47\&48}, 357 (1985).

\bibitem{Trovarelli-PRL-2000}
O. Trovarelli, C. Geibel, S. Mederle, C. Langhammer, F.M. Grosche,
P. Gegenwart, M. Lang, G. Sparn, and F. Steglich, Phys. Rev. Lett.
{\bf 85}, 626 (2000).

\end{references}
\end{document}